\documentclass[final,5p,times,twocolumn]{elsarticle}
\usepackage{atlasphysics}
\usepackage{subfigure}
\usepackage{mathrsfs}
\usepackage{graphicx}
\usepackage{hyperref}
\usepackage{multirow}
\usepackage{rotating}
\usepackage{url}
\usepackage{textpos}
\usepackage{color}
\usepackage{epstopdf}

\biboptions{sort,merge}

\newcommand{\monojettitle}{Search for new phenomena with the monojet and missing transverse momentum signature 
using the ATLAS detector in $\sqrt{s}=7\TeV$ proton-proton collisions}

\newcommand{\maybesection}[1]{\par  \vspace{1.6mm} {{\bf #1}} \vspace{0.2mm}\par}

\newcommand{\ourmagptmiss}{\MET}

\newcommand{\ourpt}{p_\mathrm{T}}

\journal{Physics Letters B}

\begin{document}
\begin{frontmatter}
\title{\monojettitle}
\author{The ATLAS Collaboration}

\begin{abstract}
\begin{textblock}{6}[0,1](0,-3){\small CERN-PH-EP-2011-090}
\end{textblock}
A search for new phenomena in events featuring a high energy jet and large missing transverse 
momentum in proton-proton collisions at $\sqrt{s}=7\TeV$ is presented using
a dataset corresponding to an integrated luminosity of 
33 pb$^{-1}$ recorded with the 
ATLAS detector at the Large Hadron Collider. The number of observed events is consistent 
with the Standard Model prediction. This result is interpreted in terms of limits 
on a model of Large Extra Dimensions.
\end{abstract}
\end{frontmatter}

\maybesection{1. Introduction}

Events composed of one high transverse energy jet and large missing transverse momentum constitute 
one of the simplest and most striking signatures that can be observed at a hadron 
collider.  The main Standard Model (SM) contribution to this ``monojet'' final state is  
Z boson plus jet production where the Z boson decays to two undetected neutrinos. Processes involving
physics beyond the Standard Model have  been suggested as additional sources 
of monojet events, including 
Large Extra Dimension (LED) scenarios~\cite{paper:ADD}, 
Split Supersymmetry~\cite{ArkaniHamed:2004fb,Giudice:2004tc,Kilian:2004uj}, and 
pair production of Dark Matter
 particles in association with a jet~\cite{paper:Goodman, paper:Beltran, paper:Bai}.
In this Letter, the 
data are interpreted in the context  of a LED model.

 Large Extra Dimensions have been proposed as a way to remove the hierarchy problem~\cite{Wess:1974}  
and to explain why gravity is so much weaker than the other forces.  In the LED 
 scenario of Arkani-Hamed, Dimopoulos, 
and Dvali (ADD)~\cite{paper:ADD}, gravity propagates in the 4 + n-dimensional bulk of space-time, 
while the other SM fields are confined to our usual four dimensions. The observed large 
difference in the characteristic mass scale of gravity (Planck mass) and the electroweak 
 scale ($W$ boson mass) 
is the result of the four-dimensional interpretation of the Planck scale.
The four-dimensional Planck scale, $M_{Pl}$, is related to the 
fundamental $4+n$-dimensional Planck scale, $M_D$, by ${M_{Pl}}^2 \sim {M_D}^{2+n}R^n$, where 
$n$ and $R$ are the number and size of the extra dimensions, respectively. An appropriate 
choice of $R$ for a given $n$ allows for a value of $M_D$ close to 
the electroweak scale. The extra spatial dimensions are compactified,
 resulting in a Kaluza-Klein tower of massive graviton modes.
At hadron colliders, these graviton modes can be produced in association with a jet. 
The production processes include $qg \rightarrow qG, gg \rightarrow gG,$ and $q\bar{q}
 \rightarrow gG$, where $G$ stands for graviton, $q$ for quark, and $g$ for gluon. As 
gravitons do not interact with the detector, these processes give rise 
to a monojet signature in the final state.

Previous monojet searches performed in Run I and Run II at
 the Tevatron~\cite{D01jet:03,CDF1jet:06} found no evidence 
of physics beyond the Standard Model. 

\vspace{1.0cm}

\maybesection{2. The ATLAS Detector and Data Samples}

The ATLAS detector~\cite{Aad:2008zzm} at the LHC covers nearly the entire solid angle 
around the collision point. It consists of an inner tracking detector surrounded by a thin 
superconducting solenoid, electromagnetic and hadronic calorimeters, and an 
external muon spectrometer incorporating three large superconducting toroid magnets. 
A three-level trigger system is used to select interesting events for recording and
subsequent offline analysis. 
Only data for which all subsystems described above are 
fully operational are used. Applying these requirements to $pp$ collision data 
taken at a centre-of-mass energy of $\sqrt{s}=7\TeV$ with stable beam conditions during 
the 2010 LHC run results in a data sample with a time-integrated luminosity of
 33 pb$^{-1}$, determined with an uncertainty of 3.4\%~\cite{atlas-update-lumi-conf, atlas-lumi-pub}.

\maybesection{3. Object Reconstruction}

Jet candidates are reconstructed using the
anti-$k_t$ jet clustering algorithm~\cite{Cacciari:2008gp,*Cacciari:2005hq} with a
distance parameter of $0.4$. The inputs to this algorithm are clusters
of calorimeter cells seeded by those with energies significantly above
the measured noise. Jet momenta are constructed by performing a
four-vector sum over these cell clusters, treating each cluster as an
$(E,\vec{p})$ four-vector with zero mass.  The resulting jet energies are corrected for
the effects of calorimeter non-compensation and inhomogeneities by
using $\ourpt$- and $\eta$-dependent\footnote{ATLAS uses a right-handed coordinate system with
 its origin at the nominal interaction point (IP) in the centre of the detector
 and the $z$-axis along the beam pipe. The $x$-axis points from the IP to the centre of
 the LHC ring, and the $y$ axis points upward. Cylindrical coordinates $(r,\phi)$ are used
 in the transverse plane, $\phi$ being the azimuthal angle around the beam pipe.
 The pseudorapidity is defined in terms of the polar angle $\theta$ as $\eta=-\ln\tan(\theta/2)$.}
calibration factors based on Monte Carlo (MC) simulations and validated with extensive
test-beam and collision-data studies~\cite{atlas-jetcal}.

Electron candidates are required to have $\ourpt > 20~\GeV$ and $|\eta| <
2.47$, and to pass the {\it medium}
 electron shower shape and track selection criteria described in~\cite{Aad:2010yt}.
Muon candidates are required to have $\ourpt > 10$ \GeV{} and $|\eta| <
2.4$ and to pass the {\it combined} reconstruction criteria described in~\cite{Aad:2010yt}, 
which include the association of a stand-alone muon spectrometer
track to an inner detector track.
Muons are required to be isolated to reduce the background contribution from jet 
``punch through'' which consists of  particles originating from a high $p_T$ jet, going through  
the calorimeter and reaching the muon spectrometer. 
  The sum of the transverse momenta of the
tracks not associated with the muon in a cone of radius $R=0.2$ in $\eta - \phi$ space 
around the muon direction is required to be less than 1.8~GeV.

The measurement of the magnitude of the missing transverse momentum ($\ourmagptmiss$) 
is done using all energy deposits in the calorimeter up to $|\eta|$ of 4.5. These
clusters are calibrated taking into account the different response of the calorimeters 
to hadrons compared to electrons or photons, as well as dead material and out-of-cluster 
energy losses~\cite{met_loc_had}.

\maybesection{4. Event Selection}

Events must be accepted by an $\met$  trigger~\cite{atl:daq2011} with a nominal 
threshold of 40 GeV, evaluated using energy depositions in the calorimeters. 
The trigger is over 99\% efficient 
for events with a reconstructed $\met$ above 120 GeV. The efficiency of the 
$\met$ trigger was determined with events selected using a muon trigger.
Events are then required to pass a set of basic kinematic selections that 
are aimed at reducing electroweak,
non-collision, and detector-induced backgrounds. These selections 
require the event to have a monojet topology characterized
by one unbalanced high $p_T$ jet resulting in large $\met$.

The selections are: 
\begin{itemize}
\item Events are required to have a reconstructed primary vertex with at least five 
associated tracks.  This ensures that the recorded event is consistent with a proton-proton collision.
\item The highest $p_T$ jet is required to have a charge fraction
$f_{\small{\textrm{ch}}}=\sum p_{T}^{\small{\textrm{track,jet}}} /p_{T}^{\small{\textrm{jet}}} > 0.02$, 
where $\sum p_{T}^{\small {\textrm{track,jet}}}$ is the scalar
sum of the transverse momenta of tracks associated with the primary vertex
within a cone of radius $R=0.4$ around the jet axis, and 
$p_{T}^{\small{ \textrm{jet}}}$ is the transverse momentum as determined 
from calorimetric measurements. 
Furthermore, events are rejected if they contain any other jet
with an electromagnetic fraction\footnote{Fraction of the energy measured in
the electromagnetic calorimeter.} $f_{\small{\textrm{em}}} <0.10$, or any jet
 in the pseudorapidity range $|\eta|<2$ with
 $f_{\small{\textrm{em}}} >0.95$ and a
 charge fraction $f_{\small{\textrm{ch}}} \le 0.05$.
The requirement $f_{\small{\textrm{em}}} <0.10$ suppresses jets produced by 
cosmic rays or beam halo muons that interact in the hadronic calorimeter. 
The latter requirements reject events in which beam halo muons deposit a 
large amount of energy in the electromagnetic calorimeter while keeping a 
high efficiency for jets originating from $pp$ collisions.   
\item Additional selections to reject events with detector noise and
non-collision backgrounds are applied: 
events are rejected if any jet 
with $p_T > 20$~GeV and $|\eta| < 4.5$ does not pass all of the additional  
quality selection criteria described in Ref.~\cite{atlas-jet-clean}.

\item   Events are required to have no identified electrons or muons according to the selection criteria stated
above. Although the signal selection vetoes leptons, control regions with identified leptons
 are used in this analysis to evaluate 
the agreement between the MC predictions and the data.

\end{itemize}

Although the results of this analysis are interpreted in this Letter in terms of the LED model,
the event selections have not been tuned to maximize the
sensitivity to any 
particular theoretical model.  To maintain sensitivity to a wide range 
of models, two sets of kinematic selections, $\LowPt$ and  
$\HighPt$, are defined. 
The $\LowPt$ selections are chosen such that
the $\met$ trigger with the highest integrated luminosity is fully efficient. 
Using lower jet $p_T$ and $\met$ selections has been motivated in the past to set limits
on the pair production of Dark Matter particles~\cite{paper:Goodman}. 
The $\HighPt$ cuts are motivated by a potential increase in
sensitivity in models such as ADD where there is a benefit from reducing
the number of background events. However, enough events in the data control samples
must be left to validate the MC predictions.

The $\LowPt$ ($\HighPt$) selections are: 
\begin{itemize}
\item Highest jet $p_T > 120$~GeV and $|\eta|< 2.0$ ($p_T > 250$~GeV and $|\eta| < 2.0$). 
\item Second highest jet $p_T < 30$~GeV and $|\eta|$ $< 4.5$ ($p_T < 60$~GeV and $|\eta|$ $< 4.5$).
The threshold is raised for the $\HighPt$ region to preserve signal acceptance.   
\item $\met > 120$~GeV  ($\met > 220$~GeV).
\item  For the $\HighPt$ selection, $\Delta\phi(\textrm{jet2},\met) > 0.5$, where $\textrm{jet2}$ is the 
second highest $p_T$ jet, and the 
 third highest jet is required to have $p_T < 30$~GeV. The number of events in which 
the large value of $\met$ is caused by a mis-measurement of the second-leading jet is reduced 
by requiring a large azimuthal separation between the direction of the second-leading 
jet and the missing transverse momentum.
\end{itemize}

\newcommand\ourtwidth{2.0cm}
\newcommand\yourtwidth{1.5cm}
\newcommand\errorSTAT{(m)}
\newcommand\errorUNCORR{[u]}
\newcommand\errorUNCORRANDJES{[u+j]}
\newcommand\errorJES{[j]}
\newcommand\errorJER{(r)}
\newcommand\errorLUMI{[{$\mathcal L$}]}
\newcommand\errorACC{(a)}
\newcommand\errorLE{($\epsilon$)}
\newcommand\errorCR{(c)}
\newcommand\errorEXT{(x)}
\newcommand\errorCREXT{(c+x)}

\maybesection{5. Background Estimate and Comparison with Data}

\par

The SM background to the monojet signature is dominated by $\Znn$+jets and   
$W$+jets production, and  includes contributions from $\Zll$+jets 
($\ell = e, \mu, \tau$), multi-jet, $t\bar{t}$, 
and $\gamma$+jets processes.
The W/Z plus jets backgrounds are estimated using 
Monte Carlo event samples normalized to data in control regions. 
 The multi-jets background contribution 
is determined from data in the case of the $\LowPt$ analysis, 
while multi-jets MC simulation is employed 
for the $\HighPt$ selection. 
Potential
contributions from beam-related background and cosmic rays are
estimated using data.  The remaining
SM backgrounds from $t\bar{t}$ and $\gamma$+jets 
are determined using simulated samples. These processes,
which contribute a negligible number of events in both the $\LowPt$ region and 
the $\HighPt$ region, will not be discussed further. 

Samples of simulated $\Znn$+jets, $\Zll$+jets, and $\Wlnj$+jets events 
are 
generated using ALPGEN v2.13~\cite{Mangano:2002ea} interfaced to 
HERWIG v6.510~\cite{Corcella:2000bw} for parton shower
and fragmentation, and to JIMMY v4.31~\cite{Butterworth:1996zw} 
to model underlying event contributions. The CTEQ6L1~\cite{Pumplin:2002ea} 
parton distribution functions (PDFs) are employed,
and the cross sections are initially normalized to 
predictions calculated to next-to-next-to-leading order (NNLO) in perturbative QCD   
as determined by the FEWZ~\cite{Melnikov:2006kv} program using MSTW2008 PDFs~\cite{mstw2008}. 
These  MC predictions are subsequently normalized using control samples in data 
as detailed below.
Multi-jets
 background contributions are simulated using
LO perturbative QCD matrix elements for $2 \rightarrow 2$ processes plus parton shower 
in the leading logarithmic approximation, as implemented 
in PYTHIA v6.421~\cite{Sjostrand:2006za}, with the ATLAS-MC09 tuning that uses 
the MRST 2007 LO$^*$ PDFs~\cite{mc09tunes}. 
All the MC samples are produced using the ATLAS detector simulation
software~\cite{sim:2010} based on GEANT4~\cite{Agostinelli:2002hh}.

The SM MC predictions for the dominant W/Z plus jets background processes suffer from 
sizeable theoretical uncertainties.
These uncertainties are related
to the parton distribution functions, the renormalization and factorization 
scales in the perturbative QCD calculations, the modeling of the 
parton showers, the underlying event contributions, and the fragmentation 
of partons into hadrons. In addition, the MC predictions are affected by 
the limited knowledge of the absolute jet energy scale and resolution~\cite{jes11}, 
trigger efficiency, and the total integrated luminosity~\cite{atlas-update-lumi-conf}.
In this analysis, W/Z plus jets data control samples
are used to normalize the MC predictions, resulting 
in a significant reduction of their systematic uncertainties.

 The data control samples are selected by removing the lepton veto from the 
requirements described previously.   These samples with an identified electron or muon 
are dominated by $\Wenj$+jets and $\Wmnj$+jets events but also include 
contributions from $\Wtauj$+jets and $\Zll$+jets processes. A small contamination from $t\bar{t}$ 
production is subtracted using MC.  The normalization factors are obtained
for a given set of kinematic selections by taking the ratio of the number of
events observed in the data to the number of events predicted by the MC.  The 
kinematic selections are varied from $\met > 120$~GeV and leading jet $\pt > 120$~GeV up to
$\met > 220$~GeV and leading jet $\pt > 250$~GeV.  The $p_T$ threshold on the jet veto is 
also increased from 0 GeV (no threshold) to  the
nominal $\LowPt$ and $\HighPt$ thresholds of $30$~GeV and $60$~GeV, respectively.
The dependence of the normalization factors on the kinematic selections is discussed  
later.

 The muon candidate sample is used to normalize   
the $\Wmnj$+jets, $\Znn$+jets, and $\Zmm$+jets 
 MC predictions. To emulate these backgrounds 
where muons are not identified and leave very little energy in the calorimeters,
the $\met$ is not corrected to take into account the momentum of the 
identified muons.  For the $\LowPt$  
selections, a normalization factor of $0.99 \pm 0.06$ is obtained.  
For the $\HighPt$ region, the small number of events left after applying the selections 
results in a large 
statistical uncertainty on the normalization factor.  To reduce this uncertainty, the number of 
events in the control sample is increased by  lowering  the $\met$ and highest jet $p_T$ thresholds 
to 180 GeV and 200 GeV, respectively.  This results in a normalization factor of $0.91 \pm 0.10$.  
Since the kinematic selections are slightly lower than those of the signal region, the 
dependence of the normalization factor on the selections was studied. No significant 
dependence was observed as the $\met$, highest jet $p_T$, and jet veto thresholds were varied. 

\begin{figure}
\includegraphics[width=0.48\textwidth]{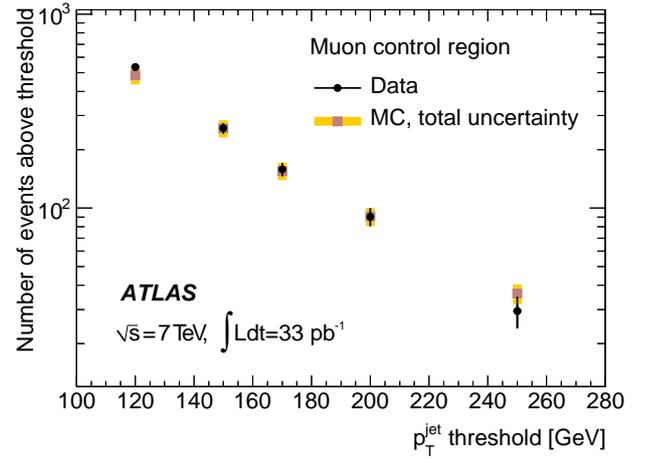}
\caption{ Observed number of events (black circles) in the muon control sample compared to the
sum of the different W/Z
plus jets predictions (squares) as a function of the highest jet $\pt$ threshold,
 in events with no second-leading jet with $\pt > 60 $~GeV.
The band indicates the total systematic uncertainty on the MC prediction.
}
\label{fig:MCscales_muo1}
\end{figure}

For the muon candidate sample, a comparison is shown in Figure~\ref{fig:MCscales_muo1} 
 between the data and the W/Z  
plus jets MC predictions normalized for the $\HighPt$ region as a function of the 
leading jet $\pt$ threshold in events with no second-leading jet with $\pt$ above 60 GeV. 

The background calculation procedure described above makes the assumption, which is supported by the 
MC simulation,  
that the normalization used for $\Wlnj$+jets backgrounds is valid for $\Zll$+jets backgrounds.
Since the largest background comes from $\Znn$+jets events and  the 
contribution from the rest of the $Z$+jets backgrounds is very small, the relevant assumption is  
that the normalization factor used for the $\Znn$+jets background should be the same as   
the one used for the $\Wmnj$+jets background.  This assumption is tested by constructing samples with 
a set of selections aimed  specifically at identifying $Z$ and $W$ bosons in events
with jets and $\met$: the $\Wmnj$ candidate events are required to have an identified muon  
with $\pt>20$ GeV, transverse mass\footnote{The transverse mass is defined as
 $m_T= \sqrt{ 2 p_{T}^{\mathrm{lepton}}   \met (1 - \textrm{cos} \Delta\phi )}$,
where $\Delta\phi$ is the azimuthal
separation between the directions of the lepton and the missing transverse momentum.}  in the region $40 < \mt < 100$~GeV, $\met > 100$~GeV, a leading 
jet with $\pt$ above 100~GeV, and no additional jets 
with $\pt > 30$~GeV. 
%  The $\met$ is corrected for the presence of the muon.
The   $\Zmm$+jets data control
samples are selected by requiring  two oppositely charged leptons with $\pt>20$~GeV, 
a dilepton invariant mass in the range 
$71 <\Mll < 111$~GeV,  $\met > 100$~GeV, a leading 
jet with $\pt$ above 100~GeV, and no additional jets 
with $\pt > 30$~GeV. The $\met$ is not corrected for the presence of the two muons.
The normalization factors are found to be $0.91 \pm 0.13$ for the $\Wmnj$ sample,
and  $0.88 \pm 0.18$ for the $\Zmm$ sample. These values are consistent 
with the normalization factors used for the background calculation.

\begin{figure}
\includegraphics[width=0.48\textwidth]{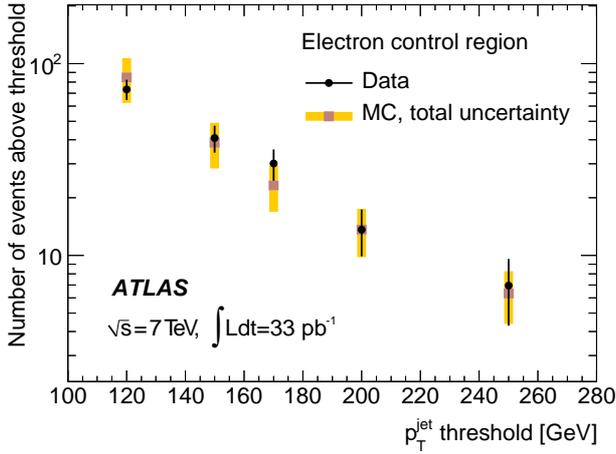}
\caption{ Observed number of events (black circles) in the electron control sample compared to the
sum of the different W/Z 
plus jets predictions (squares) as a function of the highest jet $\pt$ threshold,
 in events with no second-leading jet with $\pt > 60 $~GeV.
The band indicates the total uncertainty on the MC prediction. 
}
\label{fig:MCscales_ele2}
\end{figure}

  The electron candidate sample is used to normalize the $\Wenj$+jets, $\Zee$+jets, and 
$\Wtauj$+jets 
MC  predictions.
   Here, the electron is included in the $\met$ calculation 
since an unidentified electron can deposit a large 
amount of energy in the calorimeters.  This means that the $\met$ selection is 
qualitatively different
in the two samples even though the same value of the threshold is used. As a result,
the number of events obtained in the two samples are not expected to be the same.
  For the $\LowPt$ kinematic 
selections, a normalization factor of $0.92 \pm 0.24$ is obtained. For the $\HighPt$ 
kinematic selections, the procedure used previously for the muons of lowering the jet $p_T$ 
and $\met$ thresholds is followed and yields a 
normalization factor of $1.0 \pm 0.3$.  No significant kinematic selection 
dependence of the normalization factor was observed.  
For the electron candidate sample, a comparison in shown in Figure~\ref{fig:MCscales_ele2} 
between the data and the W/Z  
plus jets MC predictions as a function of the  
leading jet $\pt$ threshold  in events with no second-leading jet with $\pt$ above 60 GeV.

The normalization factor used for the electron backgrounds is also used for the 
 $\ensuremath{W(\rightarrow\tau\nu\rightarrow\textrm{had.+X})}$+jets  
backgrounds since the MC predicts that after all selections,   
the dominant fully hadronic $\tau$-lepton decay channel 
produces a similar reconstructed signal in the control region as that of the electron channel. The
 systematic difference in the  
 normalization factors of the two channels is much smaller than the 
uncertainties associated with the electron background normalization, and is therefore 
neglected. 
The remaining small $\Wtauj$+jets background contribution with a muon from the $\tau$-lepton decay is 
found to have a normalization factor consistent, within uncertainties,  
with the value extracted from the muon control sample. 

The total uncertainty on the electroweak background includes the uncertainties on 
the normalization factors given above, a $3\%$ uncertainty 
on the lepton identification efficiency and a $2\%$ 
uncertainty from the subtraction of the $t\bar{t}$ contribution.
Other experimental and theoretical uncertainties, that include the 
jet energy scale, jet energy resolution, luminosity, parton distribution functions, 
are cancelled, overall, by the normalization of the MC prediction to the data.

% ----  qcd 

 The multi-jets background  with large $\met$
originates mainly from the misreconstruction of the energy of the second-leading  jet 
in the calorimeters,
 resulting in  a monojet signature.
In such events, the $\met$ direction is generally aligned 
with the second-leading jet in the event. 
To estimate this background, a jets enriched data
control sample  is defined using the 
$\LowPt$ selection without the veto on the second-leading 
jet $\pt$ and requiring $\Delta \phi (\textrm{jet2},\met) < 0.5$.  Events with 
more than two jets with $\pt$ above 30~GeV are excluded. Small 
contributions from SM processes are subtracted 
according to the MC predictions. In the case of W/Z plus jets 
processes, the predictions are corrected with the 
normalization factors derived above for the relevant kinematic selections.

\begin{figure}[tb]
  \includegraphics[width=0.46\textwidth]{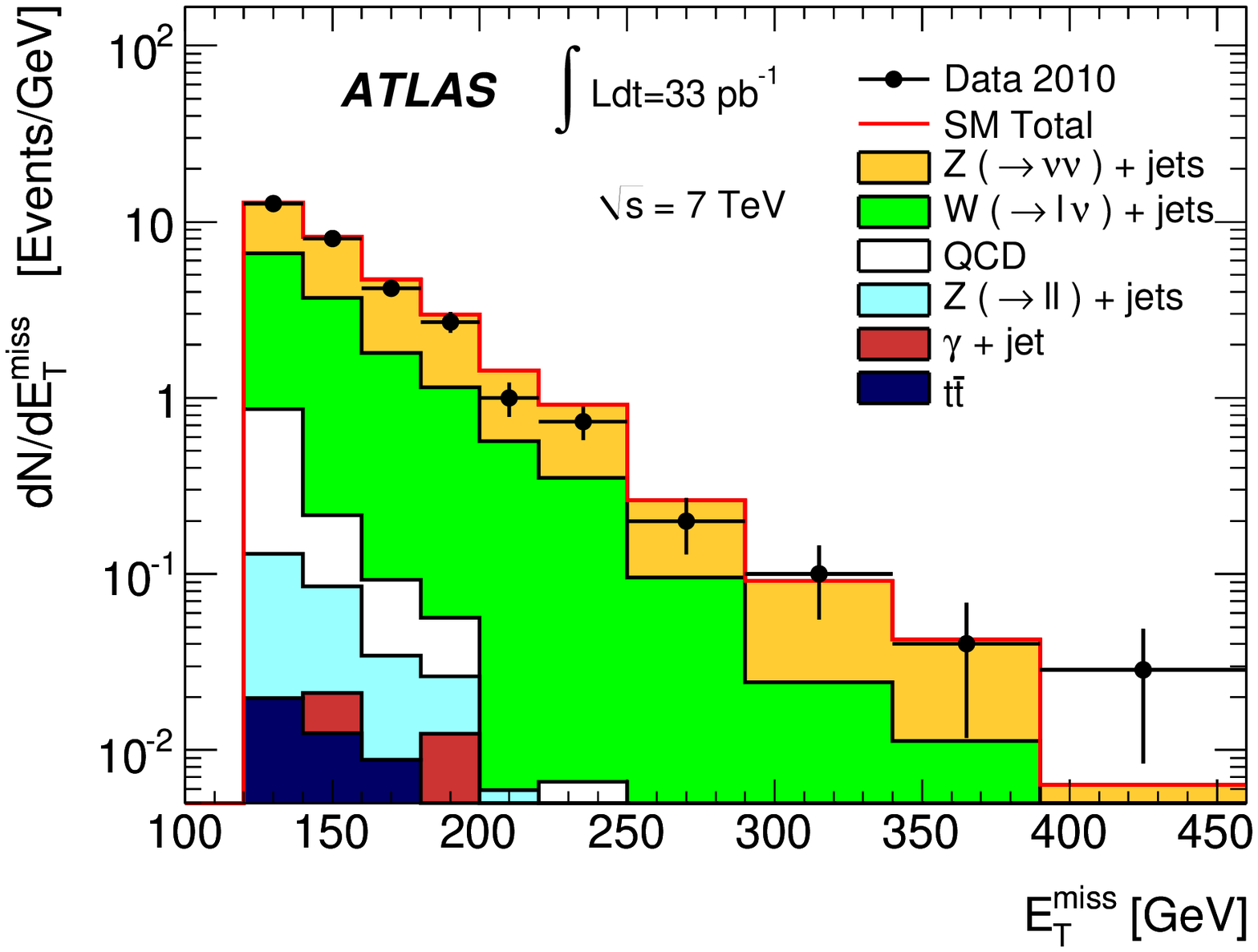}
  \includegraphics[width=0.46\textwidth]{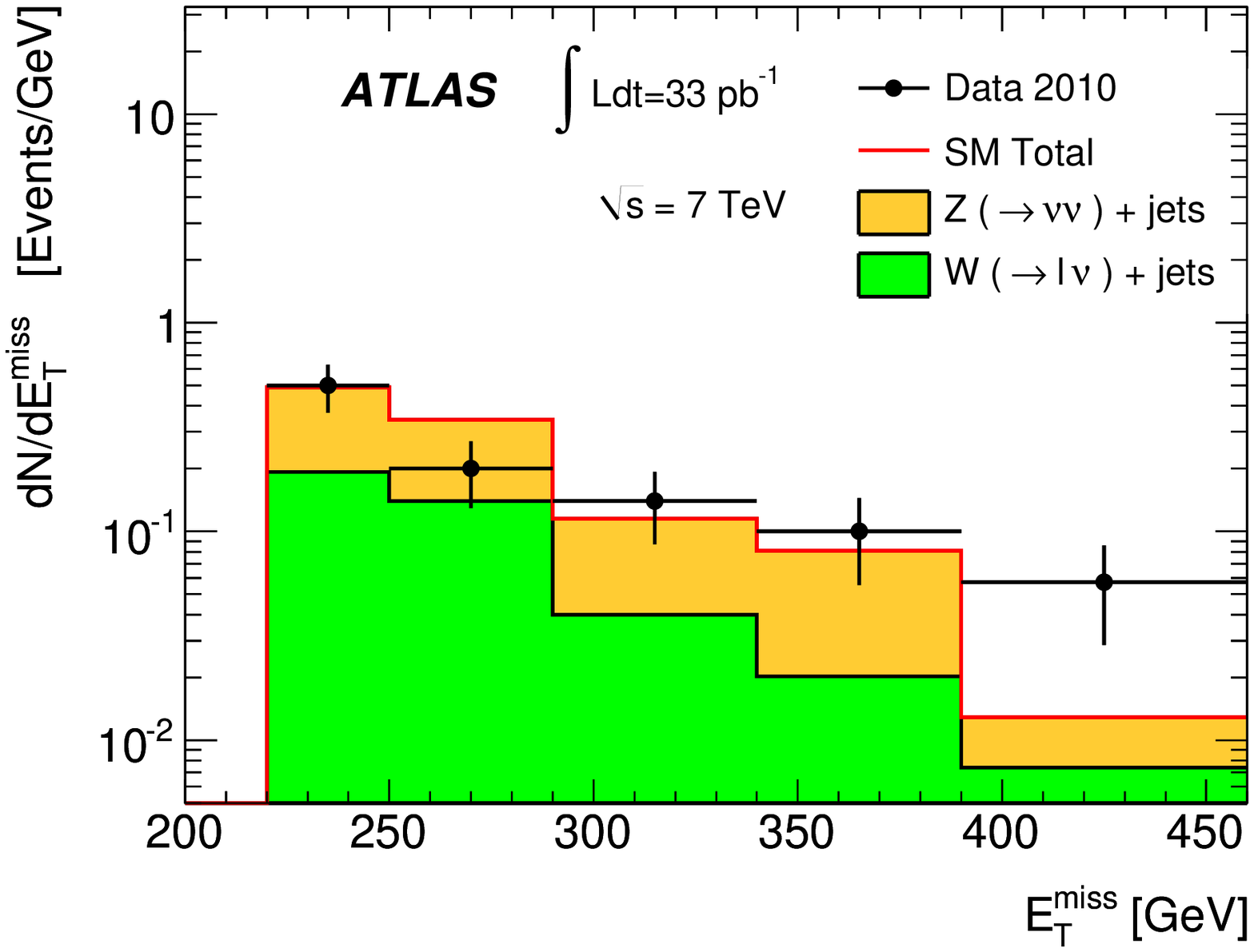}
\caption{ Measured $\met$  distributions for 
the $\LowPt$ (top) and $\HighPt$ (bottom) analyses compared to 
background predictions. Only statistical uncertainties on the data are shown. 
The systematic uncertainties on the total number of predicted events 
are 9\% for the $\LowPt$ region and 12\% for the $\HighPt$ region.}
\label{fig:MetResults}
\end{figure}

\begin{figure}[tb]
 \includegraphics[width=0.46\textwidth]{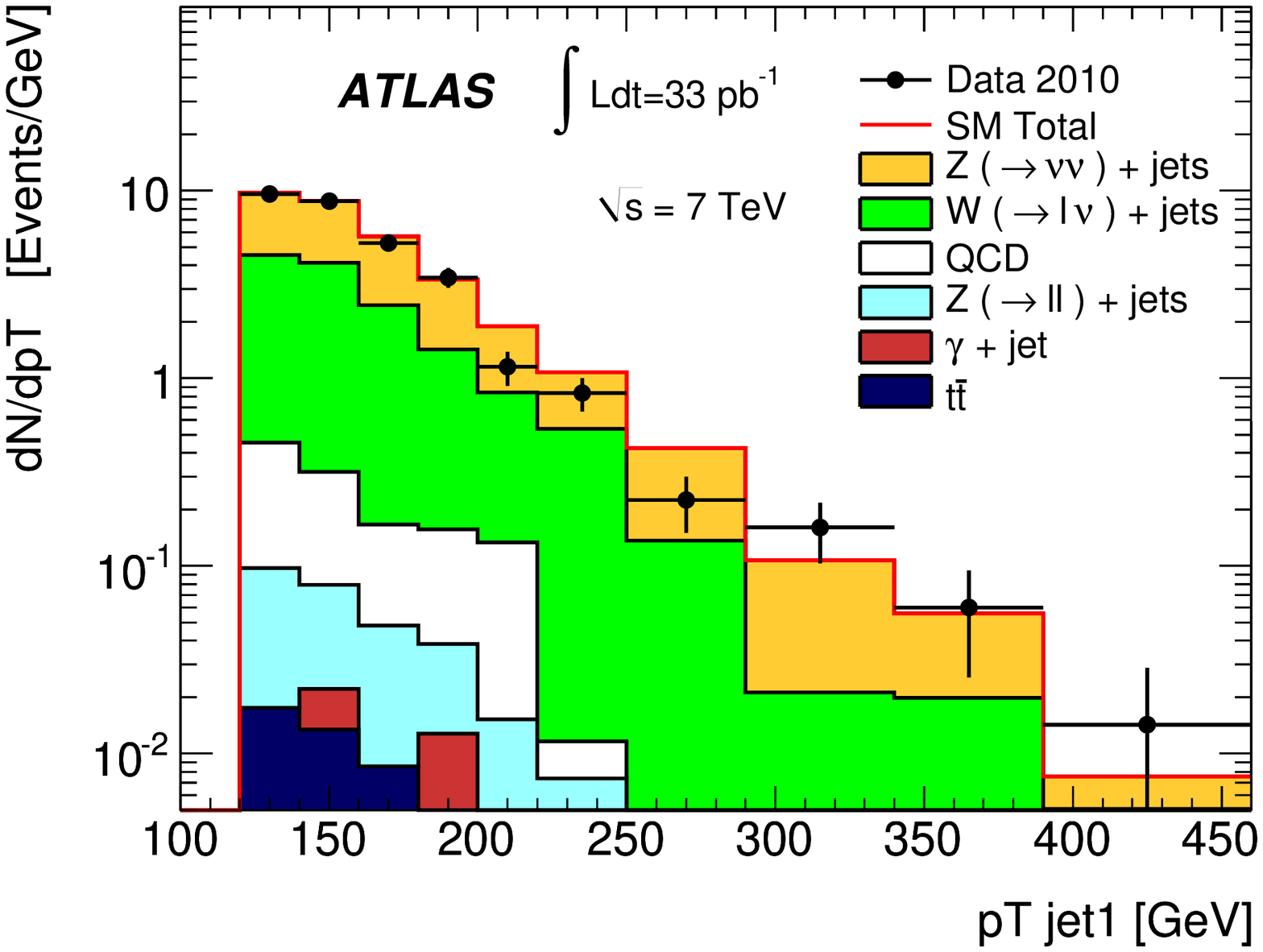}
 \includegraphics[width=0.46\textwidth]{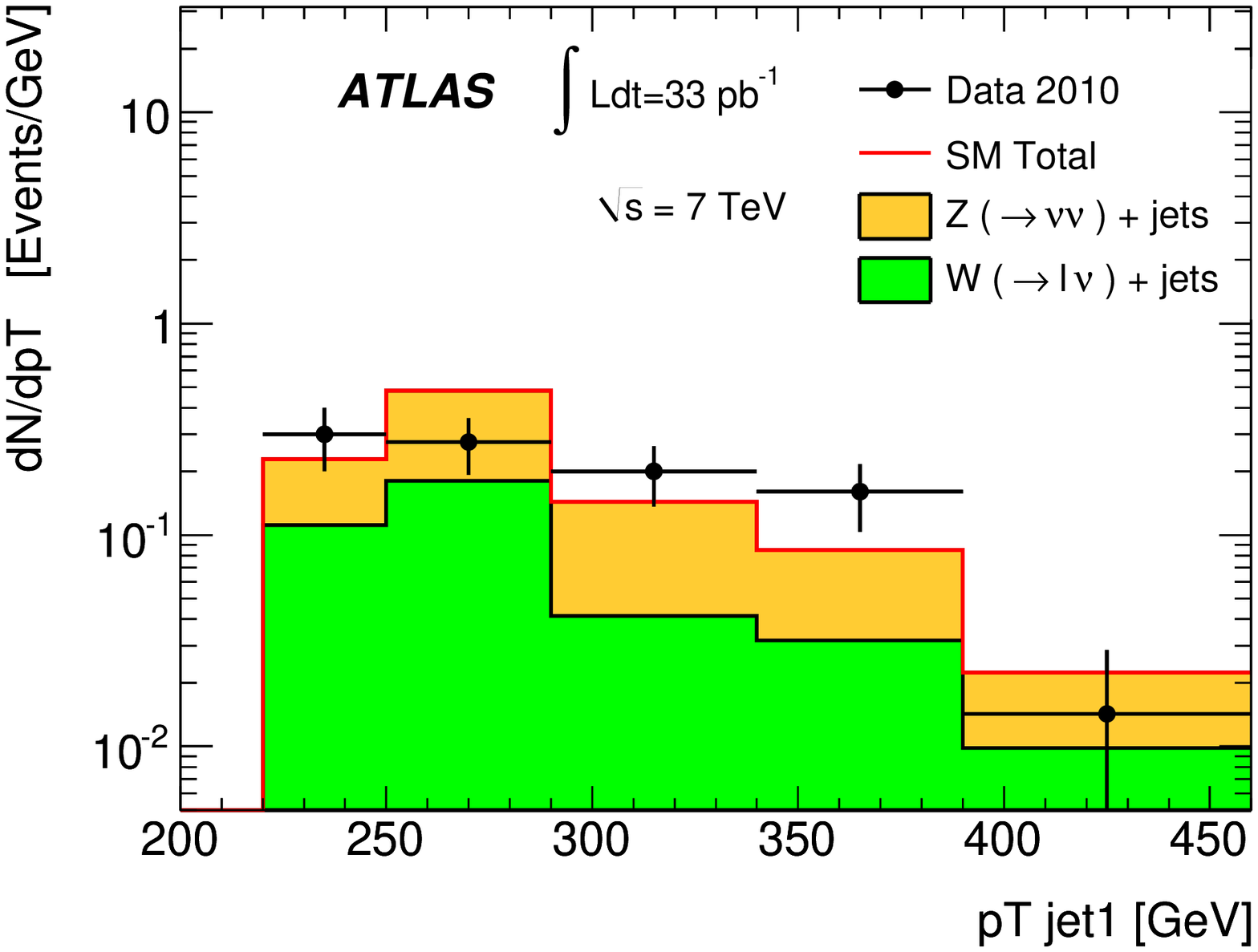}
\caption{ Measured leading-jet $\pt$ distributions for
the $\LowPt$ (top) and $\HighPt$ (bottom) analyses compared to
background predictions. Only statistical uncertainties on the data are shown.
The systematic uncertainties on the total number of predicted events 
are 9\% for the $\LowPt$ region and 12\% for the $\HighPt$ region.}
\label{fig:JetResults}
\end{figure}

The measured $\pt$ distribution of the 
second-leading jet in the jets enriched control sample is used 
to estimate the multi-jets background in the $\LowPt$ analysis. This 
estimate is  compared to PYTHIA which has to be scaled by a 
factor $1.13 \pm 0.04$
 to match the data. 
The number of multi-jets background events is obtained from a linear extrapolation 
below the threshold  of $\pt < 30$~GeV.
Several functional forms are 
considered to fit the data, and the difference with respect to the 
nominal result is included in the systematic uncertainties. 
In the $\LowPt$ analysis,  a total 
of  $24 \pm 5 \rm (stat.) \pm 14 (syst.)$ multi-jets background events are 
expected, in agreement with the $19 \pm 7 \rm (stat.)$ events predicted by PYTHIA. 
For the $\HighPt$ analysis, an estimation of the 
multi-jets background from data is not possible due to the small number of events. 
The PYTHIA MC predicts a negligible contribution.

The cosmic ray and beam-related backgrounds are estimated   
from empty  and 
unpaired proton bunches in the collider that fulfill the event 
selection criteria.
This estimate also accounts for the probability 
of  overlaps between background
contributions and  genuine proton-proton collisions leading to 
monojet signatures. A total of 
$2.4 \pm 1.1$ non-collision background events are predicted in the $\LowPt$ analysis, while the contribution in the $\HighPt$ region is negligible.

The SM background predictions are summarized in Table 1 and are found to be consistent with  
the number of observed events in the data of 611 and 39 for the $\LowPt$ and $\HighPt$ selections, respectively.
The main systematic uncertainties in the electroweak backgrounds come from 
the normalization uncertainties, which are dominated by the 
statistics in the data control samples. The 
statistical uncertainties listed in Table 1 come from the limited number of events in the MC samples.
A comparison of the SM predictions to the 
 measured $\met$ and leading-jet $\pt$ distributions are provided in 
Figures~\ref{fig:MetResults} and~\ref{fig:JetResults}, respectively.
Good agreement is observed in all cases. The results of $\chi^2$ tests performed
on the distributions of Figures~\ref{fig:MetResults} and \ref{fig:JetResults} lead to $\chi^2$ per degree of
freedom values in the range between 0.4 and 1.2.

\begin{table}[htbp]
\begin{center}
\begin{small}
\renewcommand{\baselinestretch}{1.2}
\begin{tabular}{c c c} \hline\hline
\multicolumn{3}{c}{\small{Background  Predictions $\pm$ (stat.) $\pm$ (syst.)}} \\ 
                                       & $\LowPt$ Selection &  $\HighPt$ Selection\\  \hline
$\Znn$+jets                            &$      357\pm{}12\pm{}25   $&$  25.4\pm{}2.6\pm{}2.8$\\     
$\Wtauj$+jets                           &$      139\pm{}5\pm{}36    $&$   7.8\pm{}1\pm{}2.3$\\
$\Wmnj$+jets                            &$       70\pm{}4\pm{}5     $&$   3.8\pm{}0.6\pm{}0.4$\\
$\Wenj$+jets                            &$       59\pm{}3\pm{}15     $&$   3.0\pm{}0.7\pm{}0.9$\\
Multi-jets                               &$       24\pm{}5\pm{}14    $&$-$\\
$\Ztt$+jets                            &$       2.6\pm{}0.5\pm{}0.7    $&$-$\\
$\Zmm$+jets                            &$       1.9\pm{}0.4\pm{}0.1    $&$-$\\
top                                    &$       0.96\pm{}0.04\pm{}0.2    $&$-$\\
$\gamma$+jets                          &$       0.35\pm{}{0.17}\pm{}{0.5}    $&$-$\\
$\Zee$+jets                            &$-$&$-$\\ 
Non-collision Background               &$       2.4\pm{}0.5\pm{}1.1    $&$-$\\ \hline
Total Background                       &$       657\pm{}15\pm{}62   $&$  40\pm{}2.9\pm{}4.8$\\ \hline\hline      
Events in Data  (33 pb${}^{-1}$)            &        611   &           39             \\ \hline\hline
 \end{tabular}
\end{small}
\end{center}
\renewcommand{\baselinestretch}{1}
\label{tab:Results}
\caption{
Number of observed events and predicted  background events, including statistical and
systematic uncertainties. The statistical uncertainties are due to limited MC statistics. 
The dominant systematic uncertainties come from the limited statistics in the data control regions.
The systematic uncertainties on $\Wmnj$+jets, $\Zmm$+jets, and $\Znn$+jets predictions 
are fully correlated.  Similarly, the systematic uncertainties on $\Wenj$+jets, $\Wtauj$+jets, and $\Ztt$+jets are 
fully correlated. 
}
\end{table}

\maybesection{6. Data Interpretation and Limits}

\par

\par

Since the number of events observed in the $\LowPt$ and $\HighPt$ regions are found to 
be consistent with the background predictions, as shown in Table 1, 
95\% confidence level (CL) upper limits are set on the cross-section times acceptance and 
on the value of $M_D$ as a function of the number of
extra dimensions. All limits are computed using the $CL_s$ modified frequentist approach~\cite{Junk:1999kv}. 

The 95\% CL upper limits on cross section times acceptance are calculated considering  the systematic 
uncertainties on the backgrounds and on the integrated luminosity. The resulting values are  
3.26 pb and 0.51 pb for the $\LowPt$ and $\HighPt$ analysis, respectively.

To obtain limits on the ADD parameters $M_D$ and $R$, model-dependent 
uncertainties on the signal cross sections and acceptances must be determined 
and included in the limit calculation. 

For graviton production in the ADD scenario, 
a low-energy effective field theory~\cite{paper:ADD_xsection} 
with  energy scale $M_D$ is used to calculate the  signal 
cross section considering the contribution of different graviton 
mass modes. 
Signal samples corresponding to a number of extra dimensions varying between 2 and 6 are considered, 
with the renormalization and factorization scales set to
$\frac{1}{2} M_{G}^2 + p_{T}^2$,
where $M_G$ is the graviton mass and $p_T$ 
denotes the transverse momentum of the recoiling parton.  The samples are generated 
using the  PYTHIA  MC program with
the ATLAS MC09 tuning defining all parameters including the
MRST2007 LO$^*$ PDF set.
The yields for CTEQ6.6 PDFs~\cite{Nadolsky:2008zw} are obtained by reweighting these samples. 
All generated samples are passed through the  
full detector simulation, and are reconstructed and analyzed with the same 
analysis chain as for the data.

The approximation used in the calculation of the signal cross sections is expected 
to be valid only if the scales involved in the hard interaction are significantly smaller
than $M_D$.  An estimate of the relative importance of the signal predictions in the 
unknown ultra-violet kinematic region can be made by evaluating the cross section after
rejecting events for which
 $\hat{s}  > M_D ^2$.  A substantial
contribution to the cross section from the region of phase space where $\hat{s}$  is comparable to 
or larger than $M_D^2$  would indicate that the model does not provide reliable predictions.
In the case of 2 to 4 extra dimensions, and for the $\HighPt$ selections, the change in 
the accepted cross section varies between 2\% and 28\%, respectively. 
The effect is larger when the number of dimensions is increased, and can be as large as  60\% for 6 dimensions.
In this analysis, only predictions for up to 4 extra dimensions are therefore considered.

Systematic uncertainties 
that affect the production cross section include the PDF and scale uncertainties.
The former are evaluated by studying the variations of the cross section obtained  
between the nominal CTEQ6.6 value and its 44 error sets\footnote{The errors provided by CTEQ 
correspond to 90\% confidence intervals. Here, they are rescaled
to correspond to 68\% intervals.}.
 The uncertainty on the cross section related 
to the choice of renormalization and factorization scales
is estimated by varying the scales upward and downward
by a factor of two from their nominal value.

Systematic uncertainties affecting the signal acceptance are estimated as follows.
The uncertainty coming from the modeling of initial and final state radiation (ISR/FSR)
is estimated by varying the simulation parameters 
controlling ISR and FSR within a range that is consistent with 
experimental data~\cite{skands}.
The jet energy scale (JES) and resolution (JER) are varied by their uncertainties 
(\cite{jes11} and ~\cite{atlas-jer}), and their impact on the signal acceptance is evaluated.
The contributions of the PDF and scale uncertainties to the acceptance uncertainty 
are evaluated using the methods described above.
The systematic uncertainty from the modeling of the pile-up is studied by comparing 
MC samples simulated without pile-up
and with and average of two interactions per bunch crossing, corresponding to the average
number of interactions per crossing observed in the 2010 data.

Finally, the uncertainty of 3.4\% on the luminosity is also included.
The values of all the signal-related systematic uncertainties considered above are 
given in Table~\ref{tab:systematics}, where all correlations are taken into account.

\begin{figure*}
\centering
\includegraphics[width=0.485\textwidth]{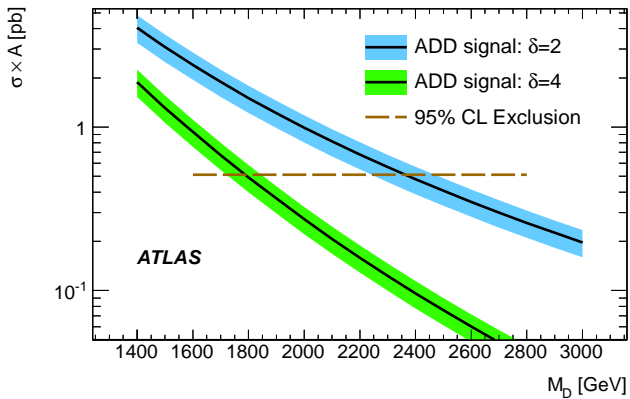}
\includegraphics[width=0.485\textwidth]{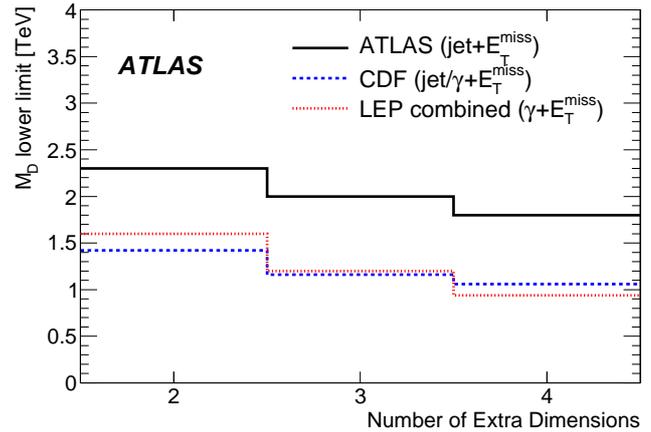}
\caption{Left: Signal cross section times acceptance as a function of $M_D$ predicted by the effective ADD theory
for 2 and 4 extra dimensions. The bands surrounding the curves reflect the systematic uncertainties.
The observed limit is shown as a dashed line. 
Right: 95\% CL observed lower limits on $M_D$ for different numbers of extra dimensions for ATLAS, 
CDF \cite{CDFadd:08}, and LEP \cite{LEPadd:04,ALEPHadd:03,DELPHIadd:05,OPALadd:2000,L3add:04}. }
\label{fig:MD_Limits}
\end{figure*}

Signal cross sections times acceptance  predicted by the effective theory for 2 and 4 extra dimensions 
are shown on the left side of Figure~\ref{fig:MD_Limits} 
as a function of $M_D$.
The bands around the curves 
reflect the uncertainties described 
previously. 
The cross section times acceptance limit of 0.51 pb is also shown for illustrative purposes.

To compute the 95\% CL limits on ADD model parameters,
all signal systematic uncertainties are considered and the 
$CL_s$ approach mentioned above is used.  Since the $\HighPt$ selections provide 
the best expected limits, they are used 
to set the observed limits.
The 95\% CL observed limits on  $M_D$
 are shown on the right side of  Figure~\ref{fig:MD_Limits}.
Table~\ref{tab:limits} lists the 95\% CL lower (upper) limits on $M_D$ ($R$), 
obtained with the CTEQ6.6 PDF set.   Using the nominal MRST PDF set, the limits are  
2.3, 2.0, and 1.8 TeV for 2, 3, and 4 extra dimensions, respectively.
The expected limits are within 5\% of the observed values.

Finally, to quantify the remaining sensitivity  of the observed limits on $M_D$ to the ultra-violet behavior 
of the theory for the different  number of extra dimensions, the  95\% CL limits on $M_D$ 
are re-calculated using  the truncated phase space region with $\hat{s} <  M_D^2$.  This translates into
no significant change in the case of 2 and 3 extra dimensions and reduces the quoted limit for 
the case of 4 extra dimensions by 10\%.

\begin{table}[htp]
\begin{center}\renewcommand\arraystretch{1.2}
\renewcommand\tabcolsep{1.2mm}
\begin{tabular}{c cc }
\hline\hline
Source & $\LowPt$ (\%)& $\HighPt$ (\%)  \\
\hline
PDFs & 6 & 7 \\ 
ISR/FSR & 13 & 13 \\ 
$Q^{2}$ scale & 11 & 11 \\ 
JES & 5 & 6 \\ 
JER & 3 & 1 \\ 
Pile-up & 3 & 2 \\ 
Luminosity & 3 & 3 \\ \hline
Total Systematics & 20 & 20 \\ \hline\hline
\end{tabular}
\caption{Systematic uncertainties (in $\%$) on ADD graviton 
signal yields for the $\LowPt$ and $\HighPt$ kinematic regions, respectively.}
\label{tab:systematics}
\end{center}
\end{table}

\begin{table}[htp]
\begin{center}\renewcommand\arraystretch{1.2}
\renewcommand\tabcolsep{1.2mm}
\begin{tabular}{ccc}
\hline \hline
$\delta$ & $M_{D}$ [TeV] & $R$ [pm]\\
\hline
      2 & 2.3 & 9.2$\times 10^{7}$ \\
	          
      3 & 2.0 & 1.1$\times 10^{3}$\\
	          
      4 & 1.8 & 4.1 \\
\hline \hline
\end{tabular}
\caption{$95\%$ CL observed lower (upper) limits on $M_{D}$ ($R$) for $\delta=2-4$,
 using a dataset corresponding to
 an integrated luminosity of 33 $\ipb$. These results 
 are obtained using the $\HighPt$ selection, and CTEQ6.6 PDF set. The expected limits are within 5\% of the observed values.} 
\label{tab:limits}
\end{center}
\end{table}

\maybesection{7. Conclusion}

A search for new physics in final states containing
a high-\pT{} jet and missing transverse momentum is performed
using  33~pb$^{-1}$ of $pp$ collision data collected by the ATLAS detector. 
Good agreement is observed between the data and Standard Model predictions
in the two kinematic regions studied in this analysis.  
95\% CL upper limits on cross section times acceptance are found   
to be 3.26 pb and 0.51 pb for the $\LowPt$ and $\HighPt$ analysis, respectively.

The results are then interpreted in terms of the ADD LED scenario
where $M_D$ values between 2.3 TeV and 1.8 TeV are 
excluded at the $95\%$ confidence level 
for a number of extra dimensions varying from 2 to 4, respectively.

%%%%%%%%%%%%%%%%%%%%%%%%%%%%%%%%%%%%%%%%%%%%%%%%%%%%%%%%%%%%%%%%%%%%%%%%%%%%%%%
% Acknowledgements
%%%%%%%%%%%%%%%%%%%%%%%%%%%%%%%%%%%%%%%%%%%%%%%%%%%%%%%%%%%%%%%%%%%%%%%%%%%%%%%
%
\maybesection{8. Acknowledgements}

We wish to thank CERN for the efficient commissioning and operation of the LHC during this initial high-energy data-taking period as well as the support staff from our institutions without whom ATLAS could not be operated efficiently.

We acknowledge the support of ANPCyT, Argentina; YerPhI, Armenia; ARC, Australia; BMWF, Austria; ANAS, Azerbaijan; SSTC, Belarus; CNPq and FAPESP, Brazil; NSERC, NRC and CFI, Canada; CERN; CONICYT, Chile; CAS, MOST and NSFC, China; COLCIENCIAS, Colombia; MSMT CR, MPO CR and VSC CR, Czech Republic; DNRF, DNSRC and Lundbeck Foundation, Denmark; ARTEMIS, European Union; IN2P3-CNRS, CEA-DSM/IRFU, France; GNAS, Georgia; BMBF, DFG, HGF, MPG and AvH Foundation, Germany; GSRT, Greece; ISF, MINERVA, GIF, DIP and Benoziyo Center, Israel; INFN, Italy; MEXT and JSPS, Japan; CNRST, Morocco; FOM and NWO, Netherlands; RCN, Norway;  MNiSW, Poland; GRICES and FCT, Portugal;  MERYS (MECTS), Romania;  MES of Russia and ROSATOM, Russian Federation; JINR; MSTD, Serbia; MSSR, Slovakia; ARRS and MVZT, Slovenia; DST/NRF, South Africa; MICINN, Spain; SRC and Wallenberg Foundation, Sweden; SER,  SNSF and Cantons of Bern and Geneva, Switzerland;  NSC, Taiwan; TAEK, Turkey; STFC, the Royal Society and Leverhulme Trust, United Kingdom; DOE and NSF, United States of America.  

The crucial computing support from all WLCG partners is acknowledged gratefully, in particular from CERN and the ATLAS Tier-1 facilities at TRIUMF (Canada), NDGF (Denmark, Norway, Sweden), CC-IN2P3 (France), KIT/GridKA (Germany), INFN-CNAF (Italy), NL-T1 (Netherlands), PIC (Spain), ASGC (Taiwan), RAL (UK) and BNL (USA) and in the Tier-2 facilities worldwide.

%%%%%%%%%%%%%%%%%%%%%%%%%%%%%%%%%%%%%%%%%%%%%%%%%%%%%%%%%%%%%%%%%%%%%%%%%%%%%%%
% Bibliography
%%%%%%%%%%%%%%%%%%%%%%%%%%%%%%%%%%%%%%%%%%%%%%%%%%%%%%%%%%%%%%%%%%%%%%%%%%%%%%

\bibliographystyle{atlasnote}
\bibliography{PUB_monojet2010}

\providecommand{\href}[2]{#2}\begingroup\raggedright\begin{thebibliography}{10}

\bibitem{paper:ADD}
{N. Arkani-Hamed, S. Dimopoulos and G.R. Dvali,}
  \href{http://dx.doi.org/10.1016/S0370-2693(98)00466-3}{Phys. Lett. {\bf B429}
  (1998)  263}.

\bibitem{ArkaniHamed:2004fb}
{N. Arkani-Hamed and S. Dimopoulos,}
\href{http://dx.doi.org/10.1088/1126-6708/2005/06/073}{JHEP {\bf 06} (2005)
  073}.
%%CITATION = HEP-TH/0405159;%%.

\bibitem{Giudice:2004tc}
{G. F. Giudice and A. Romanino,}
\href{http://dx.doi.org/10.1016/j.nuclphysb.2004.11.048}{Nucl. Phys. {\bf B699}
  (2004)  65}.
%%CITATION = HEP-PH/0406088;%%.

\bibitem{Kilian:2004uj}
{W. Kilian, T. Plehn, P. Richardson and E. Schmidt,}
\href{http://dx.doi.org/10.1140/epjc/s2004-02046-5}{Eur. Phys. J. {\bf C39}
  (2005)  229}.
%%CITATION = HEP-PH/0408088;%%.

\bibitem{paper:Goodman}
{J. Goodman, M. Ibe, A. Rajaraman, W. Shepherd, T. M. P. Tait and H. B. Yu,}
  \href{http://dx.doi.org/10.1103/PhysRevD.82.116010}{{Phys. Rev.} {\bf {D82}}
  ({2010})  {116010}}.

\bibitem{paper:Beltran}
{M. Beltran, D. Hooper, E. W. Kolb, A.C. Krusbert and T. M. P. Tait,}
  \href{http://dx.doi.org/10.1007/JHEP09(2010)037}{{JHEP} {\bf {09}} ({2010})
  {037}}.

\bibitem{paper:Bai}
{Y. Bai, P. J. Fox and R. Harnik,}
  \href{http://dx.doi.org/10.1007/JHEP12(2010)045}{{JHEP} {\bf {12}} ({2010})
  {048}}.

\bibitem{Wess:1974}
{J. Wess and B. Zumino}.
  \href{http://dx.doi.org/10.1016/0550-3213(72)90584-6}{{Phys. Lett. {\bf B49}
  (1974) 52-60}}.

\bibitem{D01jet:03}
{{D\O{}}} Collaboration,
  \href{http://dx.doi.org/{10.1103/PhysRevLett.90.251802}}{{Phys. Rev. Lett.}
  {\bf {90}} ({2003})  {251802}}.

\bibitem{CDF1jet:06}
{{CDF}} Collaboration,
  \href{http://dx.doi.org/{10.1103/PhysRevLett.97.171802}}{{Phys. Rev. Lett.}
  {\bf {97}} ({2006})  {171802}}.

\bibitem{Aad:2008zzm}
{ATLAS} Collaboration,
\href{http://dx.doi.org/10.1088/1748-0221/3/08/S08003}{JINST {\bf 3} (2008)
  S08003}.
%%CITATION = JINST,3,S08003;%%.

\bibitem{atlas-update-lumi-conf}
{ATLAS} Collaboration,
  \href{http://cdsweb.cern.ch/record/1334563}{{ATLAS-CONF-2011-011}}.

\bibitem{atlas-lumi-pub}
{ATLAS} Collaboration,
  \href{http://dx.doi.org/10.1140/epjc/s10052-011-1630-5}{{Eur. Phys. J.} {\bf
  {C71}} (2011)  {1630}}.

\bibitem{Cacciari:2008gp}
{M. Cacciari, G.P. Salam and G. Soyez,}
\href{http://dx.doi.org/10.1088/1126-6708/2008/04/063}{JHEP {\bf 04} (2008)
  063}.
%%CITATION = 0802.1189;%%.

\bibitem{Cacciari:2005hq}
M.~Cacciari et al.,
\href{http://dx.doi.org/10.1016/j.physletb.2006.08.037}{Phys. Lett. {\bf B641}
  (2006)  57}.
%%CITATION = HEP-PH/0512210;%%.

\bibitem{atlas-jetcal}
{ATLAS} Collaboration,
  \href{http://cdsweb.cern.ch/record/1281329}{{ATLAS-CONF-2010-056}}.

\bibitem{Aad:2010yt}
{ATLAS} Collaboration,
\href{http://dx.doi.org/10.1007/JHEP12(2010)060}{JHEP {\bf 12} (2010)  060}.
%%CITATION = 1010.2130;%%.

\bibitem{met_loc_had}
T.~Barillari et al.,
  \href{http://cdsweb.cern.ch/record/1112035}{ATL-LARG-PUB-2009-001}.

\bibitem{atl:daq2011}
{ATLAS} Collaboration, {CERN-PH-EP-2011-078, to be published}.

\bibitem{atlas-jet-clean}
{ATLAS} Collaboration,
  \href{http://cdsweb.cern.ch/record/1277678}{{ATLAS-CONF-2010-038}}.

\bibitem{Mangano:2002ea}
M.~L. Mangano et al.,
\href{http://dx.doi.org/10.1088/1126-6708/2003/07/001}{JHEP {\bf 07} (2003)
  001}.
%%CITATION = HEP-PH/0206293;%%.

\bibitem{Corcella:2000bw}
G.~Corcella et al.,
\href{http://dx.doi.org/10.1088/1126-6708/2001/01/010}{JHEP {\bf 01} (2001)
  010}.
%%CITATION = HEP-PH/0011363;%%.

\bibitem{Butterworth:1996zw}
J.~M. Butterworth et al.,
\href{http://dx.doi.org/10.1007/s002880050286}{Z. Phys. {\bf C72} (1996)  637}.
%%CITATION = HEP-PH/9601371;%%.

\bibitem{Pumplin:2002ea}
D.~S. J.~Pumplin et al.,
\href{http://dx.doi.org/10.1088/1126-6708/2002/07/012}{JHEP {\bf 012} (2002)
  0207}.
%%CITATION = HEP-PH/0206293;%%.

\bibitem{Melnikov:2006kv}
K.~Melnikov et al.,
\href{http://dx.doi.org/10.1103/PhysRevD.74.114017}{Phys. Rev. {\bf D74} (2006)
   114017}.
%%CITATION = HEP-PH/0609070;%%.

\bibitem{mstw2008}
{A. D. Martin, W. J. Stirling, R. S. Thorne and G. Watt,}
  \href{http://dx.doi.org/10.1140/epjc/s10052-009-1072-5}{{Eur. Phys. J.} {\bf
  {C63}} (2009)  {189}}.

\bibitem{Sjostrand:2006za}
T.~{Sj\"ostrand} et al.,
\href{http://dx.doi.org/10.1088/1126-6708/2006/05/026}{JHEP {\bf 05} (2006)
  026}.
%%CITATION = HEP-PH/0603175;%%.

\bibitem{mc09tunes}
{ATLAS} Collaboration,
  \href{http://cdsweb.cern.ch/record/1247375}{{ATL-PHYS-PUB-2010-002}}.

\bibitem{sim:2010}
{{ATLAS}} Collaboration,
  \href{http://dx.doi.org/10.1140/epjc/s10052-010-1429-9}{{Eur. Phys. J.} {\bf
  {C70}} (2010)  {823}}.

\bibitem{Agostinelli:2002hh}
{{GEANT4}} Collaboration,
  \href{http://dx.doi.org/10.1106/SO168-9002(03)01368-8}{{Nucl. Instrum. Meth.}
  {\bf {A506}} ({2003})  {250}}.

\bibitem{jes11}
{ATLAS} Collaboration,
  \href{http://cdsweb.cern.ch/record/1330713}{{ATLAS-CONF-2011-007}}.

\bibitem{Junk:1999kv}
{T. Junk,}
\href{http://dx.doi.org/10.1016/S0168-9002(99)00498-2}{Nucl. Instrum. Meth.
  {\bf A434} (1999)  435}.
%%CITATION = HEP-EX/9902006;%%.

\bibitem{paper:ADD_xsection}
{G. F. Giudice, R. Rattazzi and J. D. Wells,}
  \href{http://dx.doi.org/10.1016/S0550-3213(99)00044-9}{Nucl. Phys. {\bf B544}
  (1999)  3}.

\bibitem{Nadolsky:2008zw}
P.~M. Nadolsky et al.,
\href{http://dx.doi.org/10.1103/PhysRevD.78.013004}{Phys. Rev. {\bf D78} (2008)
   013004}.
%%CITATION = 0802.0007;%%.

\bibitem{skands}
{P. Z. Skands,} \href{http://dx.doi.org/10.1103/PhysRevD.82.074018}{Phys. Rev.
  {\bf D82} (2010)  074018}.

\bibitem{atlas-jer}
{ATLAS} Collaboration,
  \href{http://cdsweb.cern.ch/record/1281311}{{ATLAS-CONF-2010-054}}.

\bibitem{CDFadd:08}
{{CDF}} Collaboration,
  \href{http://dx.doi.org/10.1103/PhysRevLett.101.181602}{Phys. Rev. Lett. {\bf
  101} (2008)  181602}.

\bibitem{LEPadd:04}
{S. Ask, {\it in Proceedings of 32nd International Conference on High Energy
  Physics (ICHEP 2004),} World Scientific, Singapore, 2005, Vol. 2, p.1289,}.
\newblock \href{http://arxiv.org/abs/hep-ex/0410004}{{\tt
  arXiv:hep-ex/0410004}}.

\bibitem{ALEPHadd:03}
{{ALEPH}} Collaboration,
  \href{http://dx.doi.org/10.1007/epjc/s2002-01129-7}{Euro. Phys. J. {\bf C28}
  (2003)  1}.

\bibitem{DELPHIadd:05}
{{DELPHI}} Collaboration,
  \href{http://dx.doi.org/10.1040/epjc/s2004-02051-8}{Euro. Phys. J. {\bf C38}
  (2005)  395}.

\bibitem{OPALadd:2000}
{{OPAL}} Collaboration,
  \href{http://dx.doi.org/10.1040/epjc/s100520000522}{Euro. Phys. J. {\bf C18}
  (2000)  253}.

\bibitem{L3add:04}
{{L3}} Collaboration,
  \href{http://dx.doi.org/10.1016/j.physletb.2004.01.010}{Phys. Lett. {\bf
  B587} (2004)  16}.

\end{thebibliography}\endgroup

\onecolumn
\clearpage
% ATLAS Collaboration author list for 18-MAY-2011
% Data extracted on 06-Jun-2011 for paperid 92
%\documentclass[11pt]{article}
%\usepackage{a4wide}\begin{document}
\begin{flushleft}
{\Large The ATLAS Collaboration}

\bigskip

G.~Aad$^{\rm 48}$,
B.~Abbott$^{\rm 111}$,
J.~Abdallah$^{\rm 11}$,
A.A.~Abdelalim$^{\rm 49}$,
A.~Abdesselam$^{\rm 118}$,
O.~Abdinov$^{\rm 10}$,
B.~Abi$^{\rm 112}$,
M.~Abolins$^{\rm 88}$,
H.~Abramowicz$^{\rm 153}$,
H.~Abreu$^{\rm 115}$
% ...
\bigskip

% ...
$^{10}$ Institute of Physics, Azerbaijan Academy of Sciences, Baku, Azerbaijan\\
$^{11}$ Institut de F\'isica d'Altes Energies and Universitat Aut\`onoma  de Barcelona and ICREA, Barcelona, Spain\\
% ...
$^{48}$ Fakult\"{a}t f\"{u}r Mathematik und Physik, Albert-Ludwigs-Universit\"{a}t, Freiburg i.Br., Germany\\
$^{49}$ Section de Physique, Universit\'e de Gen\`eve, Geneva, Switzerland\\
% ...
$^{88}$ Department of Physics and Astronomy, Michigan State University, East Lansing MI, United States of America\\
% ...
$^{111}$ Homer L. Dodge Department of Physics and Astronomy, University of Oklahoma, Norman OK, United States of America\\
$^{112}$ Department of Physics, Oklahoma State University, Stillwater OK, United States of America\\
% ...
$^{115}$ LAL, Univ. Paris-Sud and CNRS/IN2P3, Orsay, France\\
% ...
$^{118}$ Department of Physics, Oxford University, Oxford, United Kingdom\\
% ...
$^{153}$ Raymond and Beverly Sackler School of Physics and Astronomy, Tel Aviv University, Tel Aviv, Israel\\
% ...
\end{flushleft}

%\end{document}

\end{document}